\documentclass[11pt]{article}

\usepackage{amsfonts}
\usepackage{graphicx}
\usepackage{amsmath}

\makeatletter \@addtoreset{equation}{section}

\newcommand{\be}{\begin{equation}}
\newcommand{\ee}{\end{equation}}
\newcommand{\bea}{\begin{eqnarray}}
\newcommand{\eea}{\end{eqnarray}}
\begin{document}
\title{
{\bf On      Black   Objects\\ in Type IIA Superstring Theory\\ on
Calabi-Yau Manifolds } }
\author{
\hspace*{-20pt} \small \bf Adil Belhaj\thanks{belhaj@unizar.es}   \\
{\small Centre National de l'Energie, des Sciences et des Techniques Nucleaires, CNESTEN }\\[-6pt]
{\small Cellule Sciences de la Mati\`ere, Rabat,  Morocco }\\
{\small  Groupement National de Physique des Hautes Energies, GNPHE}\\{ \small  Si\`{e}%
ge focal: FS, Rabat, Morocco }}
\maketitle \ \vskip -5in {\em \
{\rightline{GNPHE/0810} } } \vskip 4in

\begin{abstract}
The  compactification of type IIA superstring on  $n$-dimensional
Calabi-Yau manifolds is discussed. In particular,
 a conjecture is given for type IIA  extremal   $p$-dimensional black  brane attractors corresponding
  to $Ads_{p+2}\times S^{8-2n-p}$  horizon geometries.
\\[.5cm]
\end{abstract}
\clearpage
%
\section{Introduction}
Recently,  $N=2$ four-dimensional (4D)  black hole (BH) physics  has
attracted much attention and been investigated using  different
connections   with  string theory, branes  and compactifications.
One nice result in this context is the  Ooguri, Strominger and Vafa
conjecture  relating the microstates counting of 4D $N=2$  BPS BH in
type II  superstrings on Calabi-Yau (CY) threefolds  to the
topological string partition function  on the same  geometry
\cite{OSV}.  This conjecture has  been extended in \cite{Vafa1,ANV}
to open topological strings, which capture information on BPS states
on D-branes wrapped on Lagrangian submanifolds of CY 3-folds,   and
to  various toric  CY threefolds \cite{ABDS}.

On the other hand,   the 4D  $N=2$    BH  exhibits an interesting
phenomenon called the attractor mechanism \cite{FK1}-\cite{Da}
which is the topic of interest here. At the horizon, the  moduli
scalar  fields   take  fixed values  given in terms of   the  BH
charges. This   can be understood in terms of an effective potential
which depends on the charges and the moduli \cite{TT}.  Extremising
the potential with respect to the moduli, the  minimum  gives  such
fixed values.

 The 4D $N=2$ attractors  have been extended to other supergravities
   in 4D as well as to higher  dimensions \cite{Sen,SSen,L,BDSS,SS,S1,S2,BDDJSS}.
    In five dimensions,  for instance, an  M-theory realisation of such BH attractors
     has been   studied  in  \cite{L}.   The  corresponding  low-energy theory also has  $N = 2$  supersymmetry involving
real special geometry. It turns out that the 4D $N=2$ black
attractors are connected to the 5D attractors \cite{CFM}.  More
recently, the black   object attractors in 6D and 7D have been
given in \cite{BDSS}. In particular,  the extremal BPS and non BPS
black attractors in   $N = 2$ supergravity, embedded in type IIA
superstring theory and  11D M-theory on the  K3 surface,
respectively,  have been studied. The attractor mechanism equations
and their solutions have been treated  using  the criticality
condition  of the attractor potential  of such  black objects.

Here we consider type IIA superstring  on CY $n$-folds and discuss
 the corresponding  extremal $p$-dimensional  black  brane attractors.  In this scenario,
 the  attractor  near-horizon  geometries are given by products of  $Ads_{p+2}$ and $(8-2n-p)$-dimensional real spheres $S^{8-2n-p}$.
In the context of the type IIA attractor mechanism on CY $n$-folds,
we conjecture that:
\[
\begin{array}{cccc}
p=0 &   \text{Black hole charges  fix} &\to & \text{geometric moduli}\\
p=3-n &   \text{ Dyonic black  brane  charges  fix} &\to & \text{ the dilaton}\\
p\neq 0, 3-n  & \quad  \text{Black  brane charges fix} &\to  &
\text{{\bf R-R} stringy moduli.}
\end{array}
\]

\section{Type IIA superstring on  CY spaces and  black brane configurations}

We start with some general observations about type IIA superstrings
on CY  manifolds and black brane configurations in the compactified
theory. It is noted that  a  $n$-dimensional CY    manifold  is
defined by the following  conditions: Complex, K\"{a}hler, and
existence of   a global nonvanishing holomorphic $n$-form.
Equivalently, it  is a K\"{a}hler manifold with a vanishing first
Chern class  $c_1=0$ ($SU(n)$ Holonolmy group). After
compactification,  it preserves only $\frac{1}{2^{n-1}}$ of the
initial supercharges.  Note that each manifold involves   a Hodge
diagram  playing   a  crucial  role in the determination of  the
string theory spectrum  in lower dimensions \cite{CHSW,G,As}. For
later reference, we table  the Hodge diagrams for    $n=1,2,3$
corresponding to $T^2$ torus, K3 surface and  CY threefolds,
respectively:
\[
\begin{tabular}{|l|l|l|}
\hline
$n=1$ & $%
\begin{tabular}{lll}
& $h^{0,0}$ &  \\
$h^{1,0}$ &  & $h^{0,1}$ \\
& $h^{1,1}$ &
\end{tabular}%
$ & $%
\begin{tabular}{lll}
& $1$ &  \\
$1$ &  & $1$ \\
& $1$ &
\end{tabular}%
$ \\ \hline
$n=2$ & $%
\begin{tabular}{lllll}
&  & $h^{0,0}$ &  &  \\
& $h^{1,0}$ &  & $h^{0,1}$ &  \\
$h^{2,0}$ &  & $h^{1,1}$ &  & $h^{0,2}$ \\
& $h^{2,1}$ &  & $h^{1,2}$ &  \\
&  & $h^{2,2}$ &  &
\end{tabular}%
$ & $%
\begin{tabular}{lllll}
&  & $1$ &  &  \\
& $0$ &  & $0$ &  \\
$1$ &  & $20$ &  & $1$ \\
& $0$ &  & $0$ &  \\
&  & $1$ &  &
\end{tabular}%
$ \\ \hline
$n=3$ & $%
\begin{tabular}{lllllll}
&  &  & $h^{0,0}$ &  &  &  \\
&  & $h^{1,0}$ &  & $h^{0,1}$ &  &  \\
& $h^{2,0}$ &  & $h^{1,1}$ &  & $h^{0,2}$ &  \\
$h^{3,0}$ &  & $h^{2,1}$ &  & $h^{1,2}$ &  & $h^{0,3}$ \\
& $h^{3,1}$ &  & $h^{2,2}$ &  & $h^{1,3}$ &  \\
&  & $h^{3,2}$ &  & $h^{2,3}$ &  &  \\
&  &  & $h^{3,3}$ &  &  &
\end{tabular}%
$ & $%
\begin{tabular}{lllllll}
&  &  & $h^{0,0}$ &  &  &  \\
&  & $0$ &  & $0$ &  &  \\
& $0$ &  & $h^{1,1}$ &  & $0$ &  \\
$1$ &  & $h^{2,1}$ &  & $h^{1,2}$ &  & $1$ \\
& $0$ &  & $h^{1,1}$ &  & $0$ &  \\
&  & $0$ &  & $0$ &  &  \\
&  &  & $1$ &  &  &
\end{tabular}%
$ \\ \hline
\end{tabular}%
\]
{}From this table, we   can  see that each   diagram  contains  two
central   orthogonal lines, obtained by deleting  the zeros.  The
vertical line  represents  parameters of the K\"{a}hler deformations
and the horizontal one  encodes the moduli of the complex structure
of the CY  $n$-folds.  The number of parameters representing the
K\"{a}hler  deformations of the metric  is  fixed  by $h^{1,1}$;
while  the number of the  complex structure deformations is given by
$h^{n-1,1}$.  The mirror version is obtained by interchanging  the
vertical and horizontal lines.

\def\m#1{\makebox[10pt]{$#1$}}
It turns out that the Hodge diagram of a CY  manifold    carries not
only geometric information (K\"{a}hler and complex deformations) but
also physical information. Indeed,  the moduli space of CY  type IIA
superstring  compactification  involves  three different
contributions depending on the entries of the Hodge diagrams. They
are given by:
\begin{itemize}
\item The geometric  deformations  of the  CY  space  including the antisymmetric B-field of the
{\bf NS-NS} sector.  This involves    the  complex structure
deformations, the   complexified  K\"{a}hler  deformations   or
both.
\item   The dilaton   defining the  string coupling  constant.
\item The scalar  moduli coming from  {\bf R-R}   gauge   fields  on non trivial cycles of the CY  spaces.
\end{itemize}
 Note that in the  case of the  K3 surface, the last  contribution   does not appear due the fact that $h^{1,0}=h^{1,2}= h^{0,1}=h^{2,1}=0$.  Then  we have only two parts.    Indeed,  the 10D perturbative bosonic  massless sector reads
\begin{equation}
\text{\bf NS-NS}: g_{MN},\;\; B_{MN}, \;\; \phi \qquad \text{\bf
R-R}: A_M, \;\;C_{MNK}
\end{equation}
where $M,N,K = 0,\ldots,9$.  The  K3 surface compactification  gives
$N=2$ supergravity  in six dimensions \cite{As}. Its spectrum
reads
\begin{equation}
g_{\mu\nu},\;\; B_{\mu\nu}, \;\; \phi, \;\;  A_\mu,
C_{\mu\nu\rho},\;\;  C_{\mu ij}, X_s.
\end{equation}
   $ g_{\mu\nu}$ is the 6D metric, $B_{\mu\nu}$  and $C_{\mu\nu\rho}$  are the 6D antisymmetric gauge fields, $A_\mu$
is  6D gravi-photon and $ C_{\mu ij}$ are  Maxwell gauge fields
coming     from the compactification of $C_{\mu\nu\rho}$  on the
real 2-cycles of the K3 surface.  Since $C_{\mu\nu\rho}$ is dual to
a vector in six dimensions, the theory has an $U(1)^{ 24}$  abelian
gauge symmetry. In addition to the 6D  dilaton  $\phi$,   there are
$80$  scalar fields  $ X_s, \;s=1,\ldots,80$. The total moduli
space of type IIA superstring on the  K3 surface is given by
\begin{equation}
\label{modulik3} \frac{SO(4,20)}{SO(4)\times SO(20)}\times SO(1,1).
\end{equation}
The first factor $\frac{SO(4,20)}{SO(4)\times SO(20)}$  determines
the geometric  deformations  of  the  K3  surfac in the presence of
the antisymmetric B-field of the  {\bf NS-NS} sector, while
$SO(1,1)$ corresponds to the dilaton scalar field \cite{BDSS}.
 This  factorization is  related  to the existence of  two black object  solutions in six dimensions,
 required by the electric/magnetic duality  represented  by the condition  \cite{BDSS}
\begin{equation}
p+q=2
\end{equation}
This condition can be solved by  two  black object configurations
\[
\begin{array}{ccc}
p=0 &  \quad  \text{BH with near-horizon geometry } & Ads_{2}\times  S^{4}\\
p=1 & \quad  \text{black string with near-horizon geometry} &
Ads_{3} \times S^{3}.
\end{array}
\]
Note that  $p=2$,  which is dual to $p=0$,  corresponds to a black
2-brane with  $Ads_{4} \times S^{4}$ near-horizon geometry. In six
dimensions, we have the following properties:
\begin{itemize}
\item $\frac{SO(4,20)}{SO(4) \times SO(20) }$  corresponds to the BH   with  24  charges
 which  is exactly  the entries appearing in  the K3  Hodge diagram.  Its  brane  realization
   is  given by the following  Hodge diagram configuration
\begin{equation}
  {\arraycolsep=2pt
  \begin{array}{*{5}{c}}
    &&\m1&& \\ &&&& \\ \m1&&\m{20}&&\m{1} \\
    &&&& \\ &&\m1&&
  \end{array}} \;=\;{\arraycolsep=2pt
  \begin{array}{*{5}{c}}
    && D0&& \\ &&&& \\ D2&&D2&&D2\\
    &&&& \\ &&D4&&
  \end{array}}
\end{equation}
Using  the electric-magnetic duality, this  factor  describes also
the moduli space of the  black 2-brane,   realized by the following
Hodge brane configurations
\begin{equation}
  {\arraycolsep=2pt
  \begin{array}{*{5}{c}}
    &&\m1&& \\ &&&& \\ \m1&&\m{20}&&\m{1} \\
    &&&& \\ &&\m1&&
  \end{array}} \;=\;{\arraycolsep=2pt
  \begin{array}{*{5}{c}}
    && D2&& \\ &&&& \\ D4&&D4&&D4\\
    &&&& \\ &&D6&&
  \end{array}}
\end{equation}

\item $ SO(1,1)$ corresponds  to  the black string  solution  with brane charges
\begin{equation}
  {\arraycolsep=2pt
  \begin{array}{*{5}{c}}
    &&\m1&& \\ &&&& \\ \m1&&\m{20}&&\m{1} \\
    &&&& \\ &&\m1&&
  \end{array}} \;=\;{\arraycolsep=2pt
  \begin{array}{*{5}{c}}
    && F1&& \\ &&&& \\ &.&.&.&\\
    &&&& \\ &&NS5&&
  \end{array}}
\end{equation}
\end{itemize}
 Having discussed the compactification of type IIA superstrings on CY manifolds,
  in what follows  we will  present a   conjecture concerning the  corresponding attractor
    mechanism  for  extremal black  branes.

\section{Conjecture on type IIA   extremal black  branes   on CY manifolds}

Before  going ahead, let us recall some results  obtained in this
context. Consider type IIA superstring   propagating   on a   CY
3-folds. This  leads to a   4D $N=2$ supergravity   coupled to
$U(1)^{h_{1,1}}$  abelian gauge symmetry.  In this theory, the
scalars    belonging  to the  vector multiplets are  given by the
complexified  K\"{a}hler  moduli space.  The  complex structure
deformations correspond to  3-cycles,  and the {\bf R-R}  gauge
fields     contribute   by  scalars in the hypermultiplets. However,
this sector  has no role in the study of  the  4D BH attractors.
Indeed,  several   studies of the  4D  type IIA  super attractor
mechanism   reveal  that in  the near horizon geometry only  the
complexified K\"{a}hler moduli can be fixed by the abelian BH
charges \cite{FK1}-\cite{Da}.  This can be obtained by minimizing
the  scalar  effective potential of the 4D BH \cite{TT}. This
potential  appears in the action of the  $N=2$ supergravity theory
coupled  to the  Maxwell  theory in which the abelian  gauge vectors
come  from the reduction of   the antisymmetric 3-form  on 2-cycles
of   CY 3-folds\footnote{Note that   the parameters of the complex
structure has been fixed in the context of type IIB superstrings
using mirror symmetry.}. The other  moduli, corresponding  either to
the   complex structure  deformations or to the moduli coming from
the {\bf R-R} gauge fields on CY cycles, remain free and  take
arbitrary  values in the near horizon limit of BH.   We believe that
this is  due to  the absence of higher-dimensional  black objects
in  4D\footnote{This  can be seen from the electric-magnetic duality
equation.}. However, the situation in six dimensions is somewhat
different. It is obtained by a compctification on  the  K3 complex
surface. The latter has a mixed  geometric moduli space involving
both   the complexified K\"{a}hler and  the complex structure
deformations.   The corresponding attractor mechanism for the black
objects has been studied in  \cite{BDSS}, see also \cite{S1,S2,SS}.
Indeed, using a matrix formulation,  the  geometric moduli of the K3
surface including the {\bf NS-NS} B-field  are fixed by the  abelian
BH charges. However,  the   dilaton  has been  fixed  by   the
black fundamental string charges (the electric and magnetic
charges).

 Having discussed  the known  results,   we now   present
    our  conjecture  for the CY attractor mechanism.   The  main  idea is
    that  the moduli of   type IIA  superstrings on a CY manifold   could  be fixed by
      the  charges of   extremal  black  branes appearing in  the compactified theory.
      Indeed, consider type IIA superstrings on a  $n$-dimensional CY manifold. This
       compactification gives  supergravity models with $ 2^{6-n}$ supercharges coupled to an
        abelian gauge symmetry\footnote{Notice that the non abelian gauge theories could  be obtained from the singular limit of such geometries.}.  The near horizon  of the corresponding black objects is defined   by the product of  Ads spaces and  spheres:
\begin{equation}
Ads_{p+2}\times S^{8-2n-p}.
\end{equation}
The integers  $n$ and $p$  satisfy
\begin{equation}
2 \leq 8-2n-p
\end{equation} The electric/magnetic duality relating
 the $p$-dimensional   electrical black   branes  to
$q$-dimensional magnetic ones reads
\begin{equation}
\label{gd} p+q=6-2n.
\end{equation}
This equation  can be solved in different  ways, here indicated by
the values for $(p,q)$:
\begin{itemize}
\item $(p,q)=(0,6-2n)$; describing an electrical charged  BH.
\item $(p,q)=(3-n,3-n)$; describing  Dyonic black branes.
\item $(p,q)\neq(0,6-2n)$ and $(p,q)\neq (3-n,3-n)$; describing black objects  like strings, membranes and
higher-dimensional branes.
\end{itemize} More precisely,  we conjecture  the following:
\begin{itemize}
\item The BH charges     fix  only the geometric  deformations  of the  CY  space  including the
{\bf NS-NS} B-field.  This involves    the  complex structure
deformations, the   complexified  K\"{a}hler  deformations   or
both.
\[
\begin{array}{ccc}
p=0 &  \quad  \text{Black hole charges  fix} \to & \text{geometric
moduli}
\end{array}
\]
\item   The dilaton   could be fixed by   the dyonic  object.
\[
\begin{array}{ccc}
p=3-n &  \quad  \text{ Dyonic black  brane  charges  fix} \to &
\text{ the dilaton}
\end{array}
\]
\item The moduli coming from {\bf  R-R}   gauge    fields on CY cycles   should   be fixed by the
higher-dimensional black   object   charges, like  strings and
branes.
\[
\begin{array}{ccc}
p\neq 0, 3-n  & \quad  \text{Black  brane charges fix} \to  &
\text{{\bf R-R} stringy moduli.}
\end{array}
\]
\end{itemize}

{\bf Acknowledgments.}   The author   would like to thank  R. Ahl
Laamara, M. Asorey,  B. Belhorma,
  L. J.  Boya, J. L. Cortes,  P.  Diaz,  L. B. Drissi,  H. Jehjouh, S. Montanez,    J.
Rasmussen, E. H. Saidi and  A. Segui  for collaborations,
discussions
 on related subjects, correspondence and scientific helps.   He  thanks also the support by  Fisica de altas energias: Particulas, Cuerdas y Cosmologia, A/9335/07.

\end{document}